# The role of adatoms for the adsorption of F4TCNQ on Au(111)

*Richard K. Berger[1], Andreas Jeindl[1], Lukas Hörmann[1], Oliver T. Hofmann[1]\**

[1]Institute of Solid State Physics, Graz University of Technology, 8010 Graz, Austria

\* Corresponding Author: o.hofmann@tugraz.at

**Abstract**

Organic adlayers on inorganic substrates often contain adatoms, which can be incorporated within the adsorbed molecular species, forming two-dimensional metal-organic frameworks at the substrate surface. The interplay between native adatoms and adsorbed molecules significantly changes various adlayer properties such as the adsorption geometry, the bond strength between the substrate and the adsorbed species, or the work function at the interface. Here we use dispersion-corrected density functional theory to gain insight into the energetics that drive the incorporation of native adatoms within molecular adlayers based on the prototypical, experimentally well-characterized system of F4TCNQ on Au(111).
We explain the adatom-induced modifications in the adsorption geometry and the adsorption energy based on the electronic structure and charge transfer at the interface. With this, we provide deeper insight into the general mechanisms causing the incorporation of adatoms within an adlayer made of a strong acceptor molecular species.

## 1  Introduction

Metal-organic interfaces play a major role for the performance of many modern devices,[1–3] especially in the context of organic electronics. Often, thin (mono)layers of organic molecules, which are sometimes referred to as charge-injection layers, are inserted between the (metallic) electrode and the active organic material in order to improve the device performance.[2,4]. These layers serve two purposes: First, they change the effective work function of the electrode and, as a direct consequence, the level-alignment with respect to the active material.[1,5–12] Secondly, they electronically decouple the electrode and the active material. Depending (inter alia) on whether the charge-injection layer itself is metallic (i.e., exhibits density of states at the Fermi edge) or not, this can either lead to an increase in device performance[6] or a decrease due to a larger tunneling barrier [13].

It is commonly assumed that, upon the right deposition conditions, organic molecules self-assemble into ordered structures on metal surfaces. This assumption is frequently corroborated by low-energy electron diffraction (LEED) or scanning tunneling microscopy experiments, which demonstrate large domains with long-range order.[14–22] However, there are also recurrent reports where the organic does not purely self-assemble, but rather incorporates "adatoms" originating from the substrate, forming structures that are more reminiscent of 2-dimensional metal-organic frameworks.[23–31] Especially adatoms of the same species as the substrate atoms are naturally present at the surface. They can be extracted from the bulk when the adsorption energetics favors the incorporation of adatoms within the adlayer. Thus, the bulk serves as a natural adatom reservoir, and adatoms can inevitably be created in the adsorption process of the molecules. The situation is made more complicated by the fact that adatoms can often not be observed directly (by imaging methods), but are only indirectly inferred.[32] Yet, recent studies, based on the determination of adsorption heights and ab-initio calculations,[33], indicate that adatom-containing structures may be more prevalent than hitherto thought, even calling previous studies explicitly into question.[33] If this hypothesis holds true, a more detailed understanding of the role of adatoms, i.e., the impact they have on metal-organic interfaces, is urgently required.

In this work, we will focus on two suites of questions: First, why are adatoms incorporated at all? I.e., how do they affect the adsorption energetics at the interface, and how do they change the way a molecule binds to the surface? Secondly, how does the presence of adatoms affect electronic properties relevant for devices, e.g., the work function and the density of states near the interface? To answer these questions, we study the adsorption of the prototypical molecule F4TCNQ (2,3,5,6-tetrafluor-7,7,8,8-tetracyan-chinodimethane) adsorbed on Au(111) by means of dispersion-corrected density functional theory. This allows us to directly compare a situation including adatoms to a hypothetical situation were the adatoms are not present.

We find that the presence of adatoms significantly alters the charge-transfer between substrate and adsorbate. While the overall charge of the molecule remains reasonably similar (approx. neutral with adatom versus 0.2 e$^-$ net charge on F4TCNQ), both the charge backdonation from the cyano groups and the charge donation into the LUMO become twice as large if adatoms are present. This goes along with a significant change in the binding energies. In particular the contribution from the covalent bonding (the charge backdonation) is significantly larger when adatoms are present. At the same time, we find that filling the LUMO with two electrons, which would be strongly unfavorable for many molecules, has a negligible impact here. Conversely, the geometric distortions that F4TCNQ undergoes lead to increased first and second electron affinities which offset the energetic cost of the geometric change.

## 2 Methods

All calculations were performed using the version 210413 of the FHI-aims software package.[34] The PBE[35] exchange-correlation functional was used. To account for van der Waals (vdW) interactions at the organic-inorganic interface, the method described by Tkatchenko and Scheffler[36] was employed with modified Hirshfeld parameters[37] for the Au atoms as proposed by Ruiz et al.[37] For the self-consistent field (SCF) cycles of the DFT calculations, several convergence thresholds were simultaneously employed, as recommended by best practices [38]. The change in the volume integrated root-mean square of the electron density was set to $10^{-3}$ $e^-/a_0^3$ (with $a_0$ being the Bohr radius) to reach convergence between consecutive SCF steps. The difference in the total energy for converged SCF iterations was set to $10^{-6}$ eV. Furthermore, the threshold for the sum of the eigenvalues of the Kohn-Sham states was set to $10^{-2}$ eV and forces acting on each atom were converged to $10^{-3}$ eV/Å

The FHI-aims software package provides different levels for the numerical parameters (such as the integration density) and the numerically tabulated atom centered basis functions. In this work, we use the "tight" defaults, which were shown to yield converged results in previous work.[39] Going beyond these settings, we furthermore increased the onset of the basis set cutoff potential from the default of 4 Å to 6 Å in order to obtain adsorption energies converged within 1 meV (see Supporting Information).

The reciprocal space was sampled using a Γ-centered k-grid with a k-point density (in the directions of the reciprocal lattice vectors) of approximately 14 1/Å$^{-1}$, also yielding adsorption energies that are converged to 1 meV (see Supporting Information). This corresponds to 7x12x1 k-points for the $\begin{pmatrix} 5 & 2 \\ 1 & 3 \end{pmatrix}$ unit cell (which is the experimentally determined unit cell[40]) and 5x6x1 k-points for the $\begin{pmatrix} 8 & 3 \\ 3 & 6 \end{pmatrix}$ which we use for comparison in the Supporting Information.

The Au substrate was modelled by slabs consisting of 5 layers of Au atoms, forming an Au(111) surface with the surface normal vector in z-direction. To describe the organic-inorganic interface using a three-dimensional periodic calculation, a repeated slab approach was applied. The slabs were separated by 40 Å of vacuum in the z-direction, avoiding quantum mechanical interaction between consecutive slabs. To compensate for the dipole potential jumps of asymmetric slabs, i.e. slabs that feature adsorbed molecules and/or adatoms on the top side, the built-in dipole correction of the FHI-package was applied.[41]

All adsorption geometry optimizations were performed using the trust radius method, relaxing all atoms of the molecule, the adatom and the top two layers until the maximum remaining force fell below 0.01 eV/Å.

# 3 Results and Discussion

## 3.1 Adsorption Geometry and Energy

To study the role of adatoms, we chose F4TCNQ on Au(111), because its geometric structure is well-known and experimentally confirmed by multiple independent studies.[39,40,42] Further it is one of the systems where adatoms could be clearly and unambiguously identified as part of the adlayer. The surface structure consists of only one F4TCNQ molecule and one adatom per unit cell, which reduces the computational effort and facilitates the analysis of this system (compared to larger unit cells). The F4TCNQ molecule and the adatom are arranged in an $\begin{pmatrix} 5 & 2 \\ 1 & 3 \end{pmatrix}$ Au(111) surface supercell, as shown in Figure 1.

The first step in our study is to determine the impact of the adatoms on the geometry of the interface, both laterally (i.e., how the adsorption site of the molecule changes with respect to the surface) and vertically (i.e., the adsorption height of the different atoms and the bending induced by the adsorption on the surface). The adsorption height was calculated as the vertical distance between the top layer of the substrate and the atoms of the adsorbed F4TCNQ. The top layer was defined by the average height of all Au(111) surface atoms (neglecting possible adatoms) in the reconstructed adsorption geometry. Vertical modifications in the geometry are particularly relevant, since they are experimentally accessible (e.g., via X-ray standing wave experiments [43]) that can be used to indirectly infer the presence of adatoms[33]. But also lateral positioning is important, since it gives us a first insight about how adatoms change the interaction with the surface.

Therefore, the geometry of F4TCNQ in the experimentally determined unit cell was optimized including and excluding the adatom. In these calculations, the molecule, the adatom (if present), and the top two metal layers were relaxed. The optimizations were started from four different adsorption positions, initially placing the center of the molecule in an atop, bridge, fcc or hcp hollow position.

When the adatom is present, we find two stable adsorption geometries: One where the center of the F4TCNQ is located approximately above a Au(111) bridge site and the adatom in an atop position (Figure 1a), and another one where the center of the F4TCNQ molecule is above an atop position and the adatom in a bridge site (see Supporting Information). The former is by approximately 160 meV energetically more stable (see below). Incidentally, placing the center of the molecule over the bridge position is also the energetically most favorable alignment without adatom present (see Figure 1b; other local minima are shown in the Supporting Information). We note in passing that this geometry also appears at low coverage (see Supporting Information), i.e., it is not influenced by intermolecular interactions. Furthermore, no adsorption geometry can be found where the adatom is in a hollow position, which would be its most stable position in the absence of molecules (see Supporting Information). In other words, the lateral position of the adatom is now determined by the energetically ideal position of the F4TCNQ molecule it bonds to, which itself is determined by the interaction between the molecule and the surface. This indicates that upon binding to the F4TCNQ molecules, the electronic coupling between adatom and surface is substantially weakened and the dominating interaction takes place between the adatom and the F4TCNQ molecules.

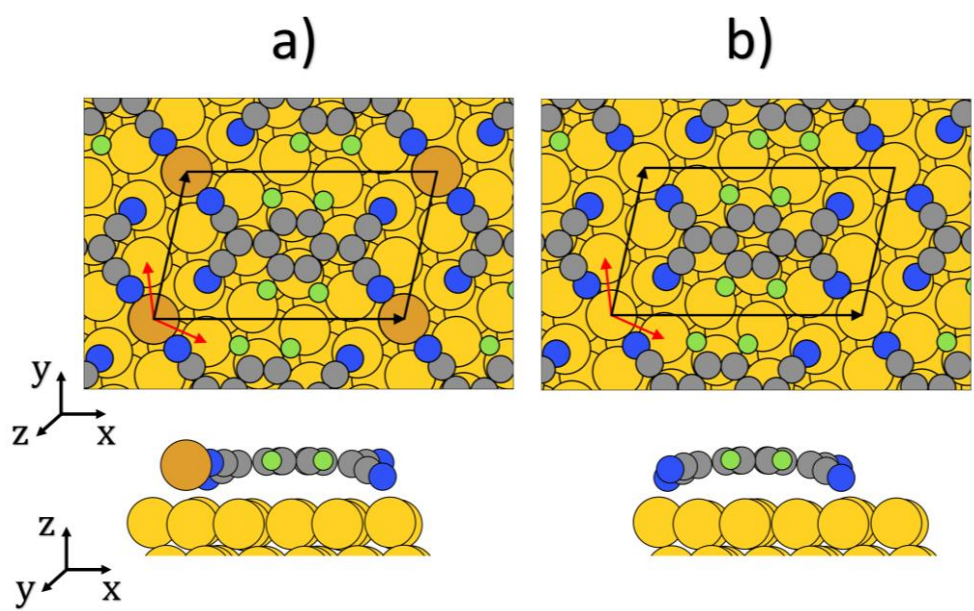

Figure 1: Minimum energy structures of F4TCNQ adsorbed on Au(111) in the $\begin{pmatrix} 5 & 2 \\ 1 & 3 \end{pmatrix}$ Au(111) surface supercell[40] (black arrows) with an Au adatom (a) and without adatoms (b). The primitive unit vectors of the Au (111) surface are illustrated by red arrows. The Au adatom is colored orange for simple discrimination to the yellow Au bulk atoms. Carbon, nitrogen, and fluorine atoms are colored grey, blue and green, respectively.

Besides the relative positioning of molecule and metal, also the adsorption heights show significant differences with and without adatoms (see Figure 2): If no adatoms are present, all four cyano groups of the F4TCNQ molecule bend towards the Au surface (Figure 2b). However, nitrogen atoms that are located at the atop sites of the Au(111) surface are more strongly bent towards the substrate than nitrogen atoms located at the hollow sites (compare Figure 1b and Figure 2b). This causes a twist of the F4TCNQ molecule in the adsorbed state.[42] Conversely, if adatoms are present, they attach to the cyano groups of two of the four neighboring F4TCNQ molecules. This causes the adatoms to be lifted from the slab towards the F4TNCQ backbone (see Figure 1a), which further corroborates the conclusion that the electronic coupling of the adatoms to the surface is weakened. The adsorption height of the adatom (3.03 Å) and the adsorption height of the backbone (3.23 Å) differ only by 0.2 Å (see Figure 2a). Consequently, the F4TCNQ cyano groups that bond to the adatom remain almost at the same height as the F4TCNQ backbone, while the cyano groups that don't bond to the adatom bend towards the substrate just like in the adatom-free case (compare Figure 2a and Figure 2b). Due to this, the twist of the F4TNQ geometry, which can already be observed for the adsorption geometry without adatoms, is further amplified by the presence of the adatom.

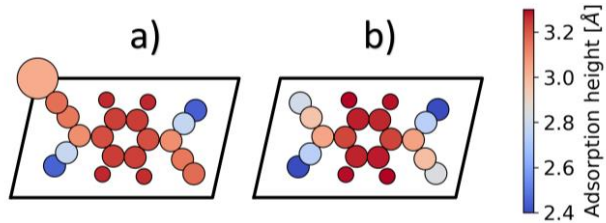

*Figure 2: Adsorption height of the geometry optimized adlayer with adatoms (a) and without adatoms (b) in the experimentally determined $\begin{pmatrix} 5 & 2 \\ 1 & 3 \end{pmatrix}$ Au(111) surface supercell [40] (black cell).*

To understand why only two of the four cyano groups remain in plane and bond to the adatom, while the other two bend downwards towards the substrate, it is instructive to analyze the nature of the bonding between F4TCNQ and the adatom in more detail. Figure 3a shows the projection of the density of states (DOS) of the full system (with adatom) onto the F4TCNQ molecule. In Figure 3b the density of states is projected onto the adatom and resolved for the different angular momenta of the wavefunction. Regions where there is significant density of states from F4TCNQ frontier orbitals are highlighted. For all frontier orbitals (i.e., the former F4TCNQ HOMO-1, HOMO and LUMO), we find that mainly the d-orbitals of the adatom contribute to the DOS. Conversely, we find contributions of the s-orbitals of the Au adatom only in regions without significant F4TCNQ density of states. This indicates a hybridization of the former F4TCNQ HOMO-1, HOMO and LUMO with mainly the d-orbitals of the Au adatom (see Figure 3).

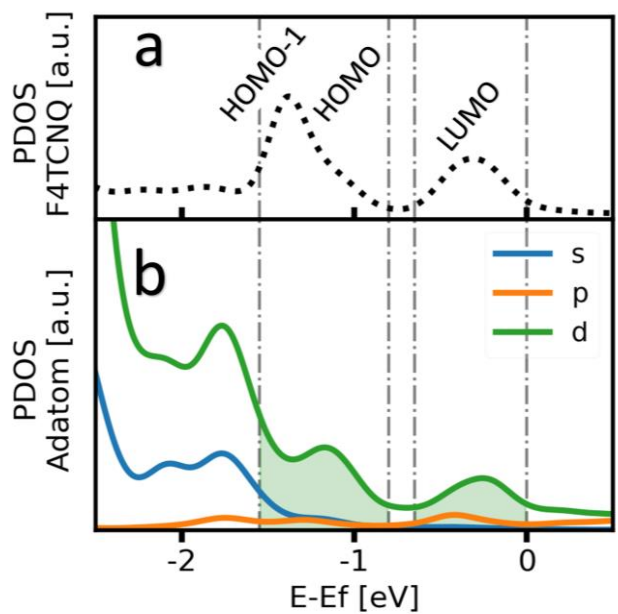

*Figure 3: a) MODOS: Total DOS projected on the MOs of F4TCNQ. b) SPDOS: Total DOS projected on all atomic orbitals of the adatom (blue) and on the angular momentum eigenfunctions of the Au adatom (l=0,1,2)*

With this, we can explain why only two cyano groups bind to the molecular adlayer: To form a bonding hybridized state, the Au atomic orbitals must match the phase of the F4TCNQ LUMO and HOMO at the cyano groups. For an Au adatom that lies within the adsorbate plane, a 5d orbital can only match this condition when it overlaps with the cyano groups of 2 neighboring F4TCNQ molecules, causing strong covalent bonding within the adlayer. Conversely, it is not possible to build any linear combination of 5d orbitals that could overlap with, for example, four neighboring F4TCNQ molecules and match the phase of the MOs at the cyano groups. This also explains why the 6s orbital of the Au adatom does not contribute to the bonding to π-orbitals: The spherically symmetrical 6s orbital cannot interfere constructively with both phases of the MOs at the cyano groups of F4TCNQ (see Supporting Information for a visualization of the relevant orbitals).

The important questions, at this point, are (i) whether the structure with an Au-adatom is indeed energetically more favorable, and (ii) if so, why that is the case, i.e., what drives the incorporation of adatoms into the F4TCNQ layer? To answer the first question, the adsorption energy ($\Delta E_{ads}$) of F4TCNQ on the Au(111) surface both with and without adatom was calculated according to Eq.: 1.

$$\Delta E_{ads} = E_{fin} - E_{init} \quad \text{Eq.: 1}$$

Eq.: 1 describes the adsorption energy as the difference between the energy of the final geometry-optimized structure after the adsorption process ($E_{fin}$) and the energy of all components in their initial geometry before the adsorption took place ($E_{init}$). Without adatoms, $E_{init}$ consists of the Au(111) slab energy of the $\begin{pmatrix} 5 & 2 \\ 1 & 3 \end{pmatrix}$ supercell and the energy of the free F4TCNQ molecule in vacuum. With adatoms, the situation is more ambiguous, as three different cases could be made: one for assuming that the adatom is already initially there, one for taking Au-atoms from kinks or step-edges, or one for taking the atom out of the bulk. Here, we opt for the third. It is, on one hand, the most conservative assumption in terms of the adatom formation energy because a bulk atom has the highest coordination number of all possibilities. On the other hand, it is also the necessary choice if the bulk is to be seen as reservoir for the Au atoms, i.e., if a whole F4TCNQ layer forms rather than just a few molecules. Based on these assumptions, we find $\Delta E_{ads}$ = -1.81 eV in the adatom-free case and $\Delta E_{ads}$ = -2.41 eV when adatoms are present, i.e., the adsorption is by -0.60 eV more favorable when adatoms are involved. This clearly shows that the phase with adatoms incorporated within the adlayer is not only kinetically trapped, but in fact, the thermodynamically stable phase.

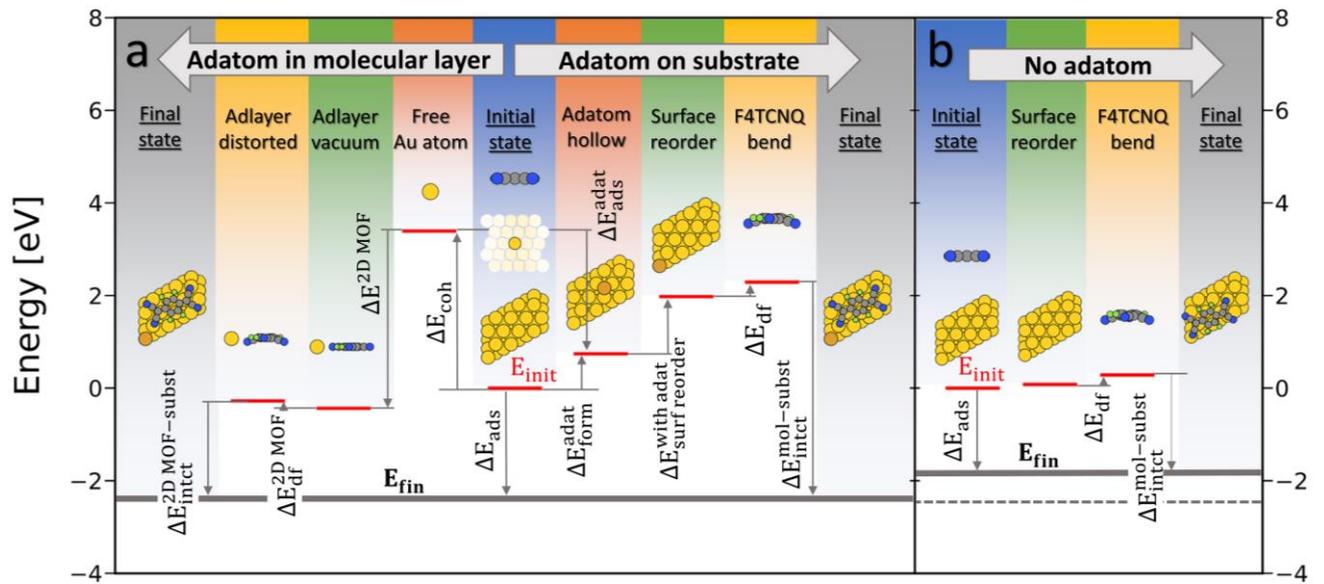

*Figure 4: Adsorption process of the F4TCNQ molecule on the Au(111) surface with adatoms (a) and without adatoms (b). The adsorption energy ($E_{ads}$) is defined as the energy difference between the final and initial state.*

At this point one may ask what makes the adatom-containing structure so much more beneficial, despite the additional cost of extracting an adatom. To shed light onto this question,, we separate the adsorption process into three hypothetical steps: 1) the "preparation" of the substrate, i.e., the energy required for the substrate atoms to rearrange into the geometry it has after adsorption. In the case of the adatom-containing structure, this includes the energy required to extract an adatom from the bulk and place it in the atop position. 2) The "preparation" of F4TCNQ, i.e., the energetic cost for a gas-phase F4TCNQ molecule to deform into the geometry it assumes on the surface. 3) The adsorption, i.e., the combination of the "prepared" geometries into the final, joint geometry. Figure 4 compares the evolution of the energy for the situation with adatom (right side of Figure 4a) and without adatom (Figure 4b).

In both cases, the energy of the "initial state", $E_{init}$, consists of the unreconstructed Au(111) slab and the energy of the free F4TCNQ molecule in vacuum ($E_{F4TCNQ}^{gasphase}$) at infinite separation (see Eq.: 2). For the case with adatom, one has to add the energy of a single Au bulk atom to the initial state energy.

$$E_{init} = E_{slab}^{unreconstructed} + E_{F4TCNQ}^{gasphase} \qquad \text{Eq.: 2}$$

As next step, the slab is set into the geometry it assumes on the surface. Without adatom, this entails only small movements of a few atoms, costing approx. $\Delta E_{prep\ slab} = 0.08$ eV (see Eq.: 4a). For the situation with adatom, conceptually first an additional Au-atom has to be taken from the bulk overcoming the cohesive energy of Au ($\Delta E_{coh} = 3.39$ eV); then, it is adsorbed in the position it would assume on its own [25] (i.e., the hollow position, $\Delta E_{adsorp}^{adat} = -2.65$ eV, see Supporting Information). Thus, the formation of an adatom (at hollow position) from the bulk amounts to an energetic cost of $\Delta E_{form}^{adat} = 0.74$ eV, according to Eq.: 3.

$$\Delta E_{form}^{adat} = \Delta E_{coh} + \Delta E_{ads}^{adat} \qquad \text{Eq.: 3}$$

Finally, all atoms of the slab are moved into the position they assume in the combined system, including moving the adatom to the atop position ($\Delta E_{surf\ reorder}^{with\ adatom} = 1.24$ eV), see Eq.: 4 b. Overall, for the adatom system ,this results in a quite significant preparation energy of the slab of $\Delta E_{prep-slab} = 1.98$ eV.

$$\Delta E_{prep-slab} = E_{slab}^{reconstructed} - E_{init} \qquad \text{Eq.: 4}$$

$$\Delta E_{surf\ reorder}^{with\ adat} = E_{slab}^{reconstructed} - (E_{slab}^{unreconstructed} + \Delta E_{form}^{adat})$$

It is important to notice that $E_{slab}^{unreconstructed}$ is the energy of the initial slab that includes the energy of one single bulk Au atom.

The F4TCNQ deformation energy ($\Delta E_{df}$) must be overcome regardless of the presence of adatoms. The deformation energy can be obtained from the energy difference between the F4TCNQ molecule in the gas geometry and the bent geometry it assumes on the surface (both calculated non-periodic), as defined in Eq.: 5.

$$\Delta E_{df} = E_{F4TCNQ}^{bend} - E_{F4TCNQ}^{gasphase} \qquad \text{Eq.: 5}$$

Naively, one could expect that the deformation without adatoms is energetically more costly, as all four CN-groups bend downwards, while otherwise two remain approximately planar. Surprisingly, we find that for the free molecule, the geometry of the adsorbed state without adatoms is the more favorable one. The deformation energy per molecule, $\Delta E_{df}$, increases from 212 meV for the adsorption geometry without adatoms, to 334 meV for the adsorption geometry with adatoms (see Figure 4 and Figure 5a). Therefore, the deformation energy change of F4TCNQ due to different adsorption geometries can be excluded as the driving force for the energetical preference of the adatom incorporating structure.

This leaves, as a last step, the combination and interaction of the two prepared subsystems. Since both the preparation of the substrate and the preparation of the molecule were energetically less favorable when an adatom is included, but the overall adsorption is more beneficial, it is immediately clear that a much more favorable interaction between the molecule and the substrate occurs when the adatom is present: To quantify this interaction we introduce the molecule-substrate interaction energy $\Delta E_{intct}^{mol-subst}$ (Eq.: 6). It is defined as the energy difference between the final structure after the adsorption and the structure made of the F4TCNQ molecule in the adsorbed adlayer geometry and the slab with the adatom at the atop position.

$$\Delta E_{intct}^{mol-subst} = E_{fin} - (E_{slab}^{reconstructed} + \Delta E_{df}) \qquad \text{Eq.: 6}$$

As Figure 4 shows, without adatom, the interaction between the molecule and the substrate is approximately $\Delta E_{intct}^{mol-subst} = -2.10\ eV$, while with adatom, it is as large as $\Delta E_{intct}^{mol-subst} = -4.72\ eV$. It is interesting to note at this point that almost the whole interaction energy without adatom is due to van der Waals interactions ($\Delta E_{vdW}$-1.76 eV), i.e., there is almost no net energy gain due to "chemical" interaction such as charge transfer and covalent bonding. Conversely, since with adatom the van der Waals energy is similarly large ($\Delta E_{vdW}$ =-1.74 eV), an energy gain of almost -2.98 eV due to charge transfer and covalent bonding is realized.

To understand to what extent the difference in interaction originates from the interaction of F4TCNQ with the adatom itself, it is useful to also look at the adsorption process in a slightly different way: The left side of Figure 4a shows a hypothetical reaction pathway where first a free Au atom is created ($\Delta E_{coh}$=3.39 eV), which then forms a 2D-MOF with F4TCNQ $\Delta E^{2D\ MOF} = -3.83$ eV and is subsequently deformed into the geometry on the surface, before it interacts with the Au slab. Notably, the interaction energy of the 2D-MOF with the surface is about the same ($\Delta E_{intct}^{2D\ MOF-subst} = -2.13$ eV) as the interaction of F4TCNQ alone with the surface and is also mostly driven by van der Waals interactions ($\Delta E_{intct}^{mol-subst} = -2.10$ eV). This is further corroborated by the fact the van der Waals interaction energy gain of the 2D-MOF when adsorbing on the Au(111) surface is only marginally smaller $\Delta E_{vdW}^{2D\ MOF}$ =-1.819 eV.

This provides a first indication that the reason for the incorporation of Au atoms into the molecular network is not driven by charge transfer with the surface, but rather due to the fact that in this geometry, a stronger covalent bond between the molecule and Au can be formed.

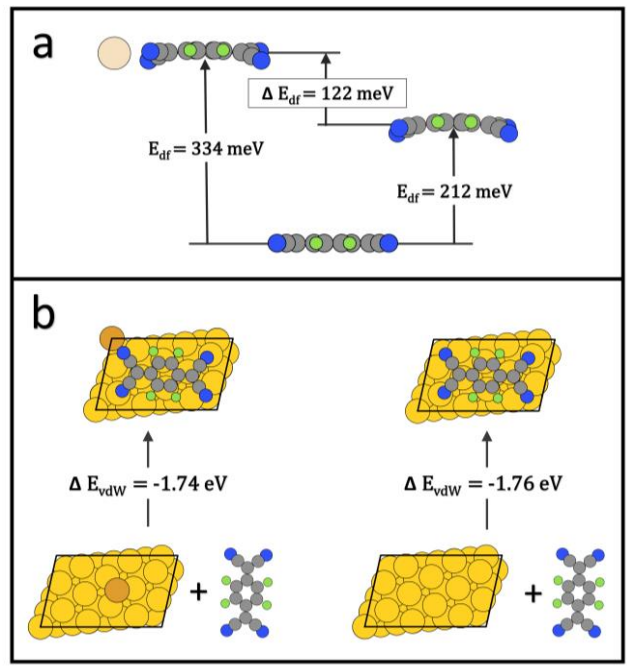

*Figure 5: Deformation energy per F4TCNQ molecule in the adlayer geometry with adatom (334 meV) and without adatom (212 meV) (a) and van-der-Waals energy of F4TCNQ in the adsorbed state described as the vdW energy difference between the slab and the complete system(b).*

A summary of the several energies used to describe the adsorption process and bond structure with and without adatoms (see Figure 4 and Figure 5) is provided in Table 1.

Table 1: Summary of all adsorption process energies introduced in the text

|  | With adatom | No adatom |
|---|---|---|
| $\Delta E_{ads}$ [eV] | -2.41 | -1.81 |
| $\Delta E_{df}$ [eV] | 0.33 | 0.21 |
| $\Delta E_{intct}^{mol-subst}$ [eV] | -4.72 | -2.10 |
| $\Delta E_{vdW}$ [eV] | -2.01 | -2.03 |
| $\Delta E_{prep\ slab}$ [eV] | 1.98 | 0.08 |
| $\Delta E_{intct}^{2D\ MOF-subst}$ [eV] | -2.13 | - |
| $\Delta E^{2D\ MOF}$ [eV] | -3.83 | - |
| $\Delta E_{coh}$ [eV] | 3.39 | - |
| $\Delta E_{ads}^{adat}$ [eV] | -2.65 | - |
| $\Delta E_{surf\ reorder}^{with\ adatom}$ [eV] | 1.24 | - |
| $\Delta E_{coh}$ [eV] | 3.39 | - |
| $\Delta E_{form}^{adat}$ [eV] | 0.74 | - |

## 3.2 Electronic structure and electrostatic potential

Overall, we have seen that most aspects (the reconstruction of the substrate, the deformation of the molecule, and even the van-der-Waals interaction between molecule and metal) are energetically worse for the situation with adatom. The only energetic contribution making the structure with adatom more favorable than the one without is the electronic interaction between the (distorted) molecules and the substrate (including adatoms). Thus, we will now shine a light on the adatom-induced changes in the electronic structure.

A useful way to determine the impact of adsorption on the electronic structure is the projection of the density of states (DOS) onto the molecular orbitals (MO), often referred to as MODOS.[44–46] This allows for determining how the MOs align energetically with respect to the Fermi level. Figure 6 compares the situation without adatom (blue) and the situation with adatom (orange). Consistent with earlier reports, we find that without adatom, the lowest unoccupied molecular orbital (LUMO) is in direct resonance with the Fermi energy,[46] while in the presence of Au-layers, the LUMO is almost completely below the Fermi energy.[39] In other words, while without adatom, F4TCNQ would assume a strongly metallic character, the presence of adatoms strongly impedes this effect.

By integrating each orbital up to the Fermi energy, the MODOS also allows insight into how the orbital occupation changes upon adsorption (Figure 6b). Most evident, and consistent with the level alignment of the LUMO in Figure 6a, Figure 6b shows the increased occupation of the former F4TCNQ LUMO in case adatoms are present in the adlayer[39] (orange dots) compared to the adatom-free case (blue crosses).

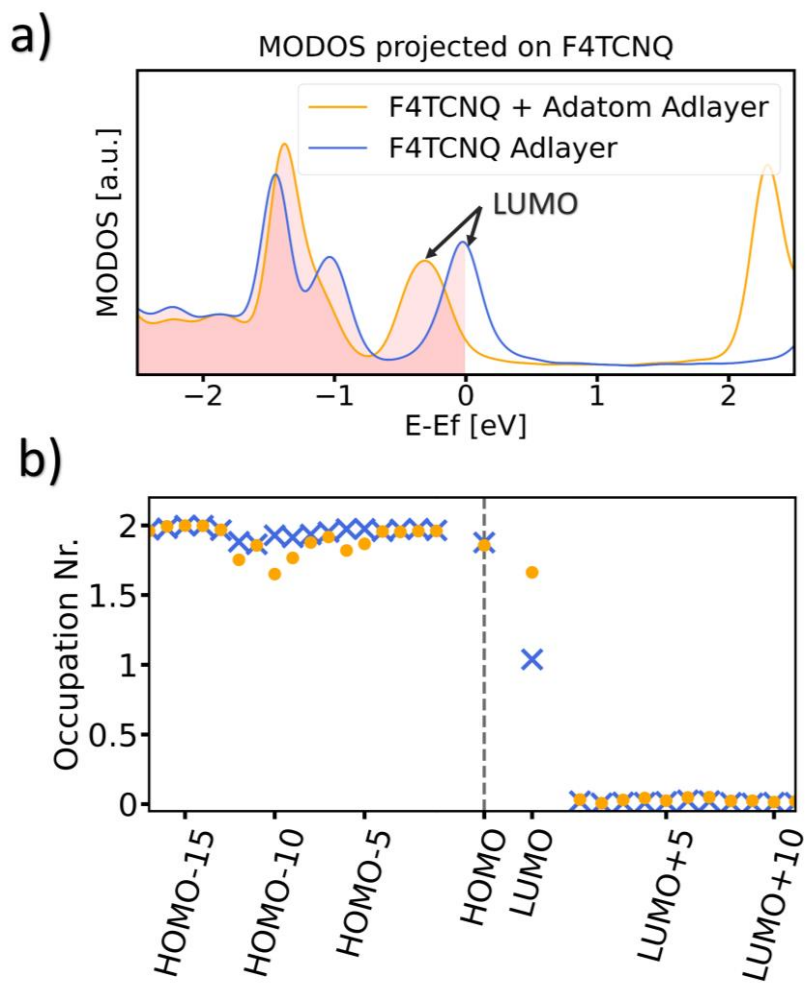

*Figure 6: a) MODOS projected on the MOs of F4TCNQ adsorbed on Au(111) with and without adatoms. b) Occupation numbers of the MOs of F4TCNQ in the adsorbed state with and without adatoms*

F4TCNQ is known to undergo a Blyholder-like charge transfer[39] with coinage metals, with a (mostly ionic) charge donation into the LUMO and a covalently driven charge backdonation originating from the molecular sigma-system[46] (particularly the cyano groups). The charge donation to F4TCNQ (i.e., forward donation) can be quantified by summing the occupation numbers, $N_i$ of all formerly unoccupied orbitals of the F4TCNQ molecule (see Eq.: 7).

$$Q_{forward} = \sum_{i=LUMO}^{\infty} N_i \qquad \text{Eq.: 7}$$

We find a forward donation of 2.2 and 1.3 electrons per adsorbed F4TCNQ molecule, with and without adatoms, respectively. Hence, the adatom causes a significant increase in the charge transferred from the Au substrate to the initially unoccupied MOs of F4TCNQ.

The vastly different occupation of the LUMO raises an important issue: For many molecules, even when the insertion of one electron into the LUMO is exothermic (i.e., the first electron affinity is negative), adding another electron to the same orbital is endothermic (i.e., the second electron affinity is positive), since for the second electron, the Coulomb-repulsion of the negatively charged molecule must be overcome. Since without adatom, approximately a single electron is added to the LUMO, while in the presence of an adatom, the LUMO gets doubly occupied (see Figure 6), it is worthwhile discussing the electron affinity of F4TCNQ. Figure 7 shows the change of the energy upon charging (i.e., the electron affinity) for a free F4TCNQ molecule in the gas phase for a) the gas-phase optimized geometry and b) the geometry it assumes on the surface with adatom. Qualitatively, as expected, for all geometries we see that the first electron affinity is strongly negative, while the second electron affinity is positive. However, the second electron affinity is already quite small for the gas phase molecule (only ~+410 meV), showing that the quantum-mechanical gain of adding an electron to the LUMO and the Coulomb-repulsion from the first electron approximately compensate. Moreover, when the molecule is calculated in its final geometry on the surface, the second electron affinities notably decrease. Qualitatively, this is to be expected, as a system out of equilibrium becomes more reactive and exhibits a smaller gap (and, thus, an energetically lower LUMO). Quantitatively, it is still interesting to notice that both the first and the second electron affinities become approx. 300 meV per electron more favorable, which is in the same order of magnitude (but of opposite sign) than the energetic cost of the deformation. In other words, adding a second electron into the F4TCNQ LUMO is, per se, associated with very little energetic cost. More importantly, however, the energetic deformation makes adding the second electron less unfavorable, and, on top of this, enhances the electron affinity by an amount that completely offsets, and even overcompensates, the energetic cost of deformation (compare Figure 5).

Besides the charge donation into the molecule, F4TCNQ simultaneously donates charge back into the substrate. This is reflected in a slight decrease in the occupation numbers of all MOs that are energetically below the F4TNCQ LUMO. According to Eq.: 8, the amount of charge shifted due to backdonation can be quantified by summing the occupation numbers of all initially occupied orbitals up to the HOMO and subsequential subtraction of the original number of electrons of the free F4TCNQ molecule (136 electrons).

$$Q_{back} = \sum_{i=1}^{HOMO} N_i - 136 \qquad \text{Eq.: 8}$$

Calculating the backdonation from the occupation numbers in Figure 6b according to Eq.: 8 yields a total charge of 2.2 and 1.1 electrons shifted towards the substrate, with and without adatoms, respectively. Overall, this means that when an adatom is present a) the charge rearrangement in both directions is significantly larger, and b) the F4TCNQ molecule is effectively charge-neutral (i.e., donation and backdonation cancel each other). In the case without adatoms, forward donation dominates over back donation, yielding a net charge transfer of 0.2 electrons per F4TCNQ molecule. Worth noting at this point is the difference in the magnitude of charge transfer in both directions. We attribute this observation to the fact that in the presence of adatoms, a much stronger covalent bond between the molecule and the metal is possible (see above). Since this bond mainly results in charge backdonation, it must be compensated by a similarly large charge donation, resulting in a more strongly filled LUMO. This corroborates the above indication why adatoms are incorporated in the first place: The ability to form stronger covalent bonds leads to an energy gain that offsets the costs of generating the adatom from the bulk.

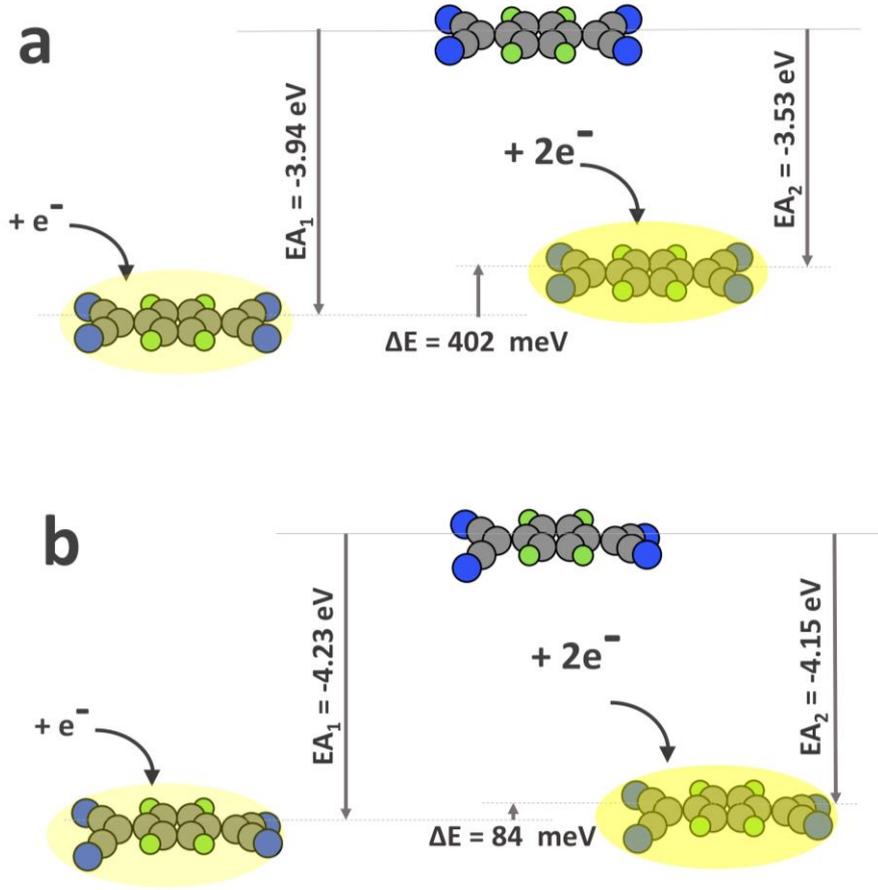

*Figure 7: Electron affinity of free F4TCNQ in the original planar geometry (a) and in the geometry of the adlayer with adatom (b)*

### 3.3 Influence of adatoms on the interface dipole

The geometry and the different charge transfer with and without adatom also have a direct impact on interface properties that are directly relevant to applications, such as the adsorption-induced work function modification. To explain this, we apply the common approach of splitting the total interface dipole, $\Delta\Phi$, into a contribution from the molecule, $\Delta E^{Mol}$, the contribution from the slab, $\Delta E^{Slab}$, and the bond dipole due to interfacial charge transfer, $BD$.[46,47] The bond dipole is caused by the total rearrangement of electron density due to the chemical interaction of the molecules with the substrate.

$$\Delta\phi = \Delta E^{Mol} + \Delta E^{Slab} + BD \qquad \text{Eq.: 9}$$

As all our calculations employ a dipole correction counteracting the potential difference between top and bottom of a geometry, the total interface dipole, as well as the geometric dipoles ($\Delta E^{Mol}$ and $\Delta E^{Slab}$) can be obtained directly from calculations of the respective subsystems. As the last constituent, the bond dipole is simply obtained via subtracting the geometric contributions from $\Delta\phi$.

The full interface dipole $\Delta\phi$ amounts to +0.23 eV without adatoms and -0.20 eV with adatoms. Its components are visualized in Figure 8. In both cases, the molecular dipole opposes the bond dipole, but due the stronger bending of the molecule geometry without adatom, also its dipole component is stronger (-0.55 eV versus -0.36 eV). In the structure with adatom, this dipole is further decreased by the substrate dipole opposing the molecular dipole (0.13 eV). For the case without adatoms, the substrate dipole can be neglected (0.01 eV)

The last contribution is the bond dipole. For the system without adatom the charge transfer from the substrate to the adlayer leads to a dipole of roughly 0.8 eV. With adatoms in the adlayer, this is reduced to only 0.03 eV due to the small net charge transfer. The bond dipole can also be calculated via the charge rearrangements of the electron density. This approach leads to slightly different results, but allows for some qualitative arguments about the origin of the dipole. Therefore, a discussion of this approach is presented in the Supporting Information.

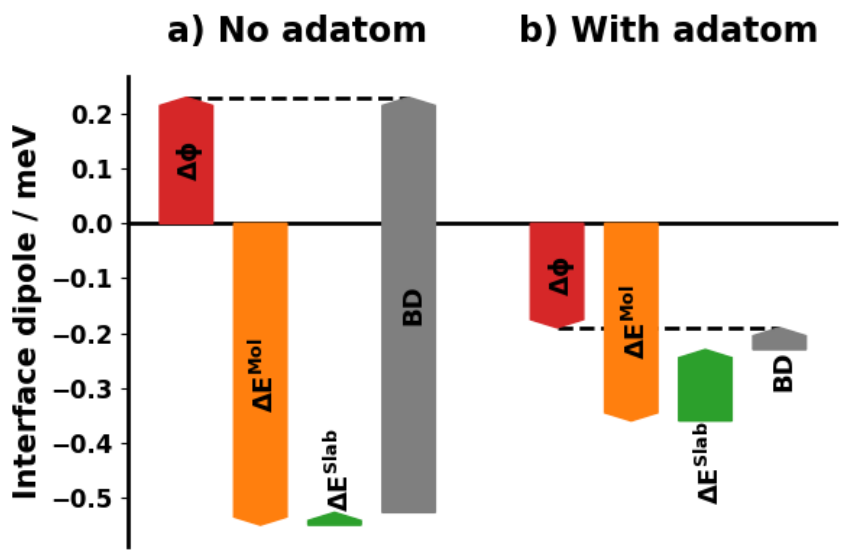

*Figure 8: Contributions to the total interface dipole $\Delta\phi$ without (left) and with adatom (right). Nature of the bond to the adatom.*

# 4 Conclusions

In this work, we discuss how surface adatoms influence the adsorption of F4TCNQ on Au(111). We find strong indications that the incorporation of adatoms is driven by the fact that the molecule, specifically its cyano groups, can form more efficient covalent bonds with adatoms than with Au atoms in the first layer. This finding is supported by the observation that the adatom is moved away from its initial equilibrium position at the Au(111) hollow site towards an Au(111) atop site (also in agreement with literature [39,42]) while being lifted up to the adsorption height of the F4TCNQ adlayer.[42] The finding is furthermore corroborated by the observation that the interaction between the surface and F4TCNQ alone, as well as with a hypothetical F4TCNQ-Au network, is driven almost exclusively by van-der-Waals forces. The energy gain from the improved (covalent) bonding is so large that it readily offsets the energetic cost of extracting an Au atom from the bulk metal.

Interestingly, despite the fact that the adatoms allow two of the four cyano groups to remain approximately in the molecular plane, the different adsorption-induced distortion of the molecule plays a minor role for the energetics. In fact, in contrast to the naïve expectation, in the presence of adatoms F4TCNQ even experiences an increased twist, which is in agreement with other reports.[42] The geometry optimization shows that the adatom only bonds to two of four neighboring F4TCNQ molecules, which can be explained by the hybridization between the atomic orbitals of the adatom and the MOs of the F4TCNQ molecule. By projecting the DOS on both the orbitals of the adatom and the molecule we find that mainly the d orbitals of the adatom hybridize with the F4TCNQ frontier MOs, allowing for a coordination of the adatom to only two of four neighboring F4TCNQ molecules.

The presence of an adatom and the ensuing stronger covalent bonding leads to a marked increase in charge backdonation, i.e., from the molecular σ-orbitals to Au, from ca. 1 electron without to 2 electrons with adatom. To maintain Fermi-level pinning, this increased backdonation is compensated by a larger donation, i.e., filling of the molecular LUMO (also here from ca. 1 electron to approx. 2 electrons). Filling the LUMO twice is energetically relatively favorable for F4TCNQ. Here, this is due to the fact that neutral F4TCNQ is a quinoid molecule (with 4n electrons in the π-system), and filling the LUMO twice thus fulfills the Hückel rule (i.e., creates a p-system with 4n+2 electrons in the π system), i.e., makes the molecule aromatic and thus particularly stable. This is a classic example for surface-induced aromatic stabilization [48].

As a tentative synopsis of these observations, we expect systems that easily sustain significantly increased charge transfer, i.e., that show a particularly low second electron affinity, to be particularly likely to extract adatoms from the bulk and include them into the molecular framework.

The presence of adatoms also directly affect various interface properties. A direct consequence of the charge transfer is that without adatoms, the LUMO is in direct resonance with the Fermi-energy, i.e., the adsorbed molecule shows a large density of states at $E_F$ ("metallicity"), while with adatoms, only the high-energy flank of the LUMO crosses $E_F$, i.e., the metallicity of the adlayer is, counterintuitively, small. Furthermore, we find that due to the modified charge transfer and molecular distortion, the adsorption induced work function modification is significantly smaller (by ca. -0.4eV) when adatom are present.

## Supporting Information

The Supporting Information contains: convergence tests for the basis function cutoff radius, convergence tests for the integration grid, explanations and visualizations of stable F4TCNQ adsorption geometries with and without adatoms, the bond dipoles calculated via electron densities, visualizations of the frontier orbitals of F4TCNQ and a gold atom.

## AUTHOR INFORMATION

**Corresponding Author**

Oliver T. Hofmann, o.hofmann@tugraz.at

**Notes**

The authors declare no competing financial interest.

**Acknowledgment**

We acknowledge inspiring conversations with Giovanni Costantini and Reinhard J. Maurer, as well as fruitful discussions with F. Calcinelli, J. J. Cartus, A. Werkovits, and B. Ramsauer. Funding through the START project of the Austrian Science Fund (FWF): Y1157-N36 is gratefully acknowledged. Computational results have been achieved using the Vienna Scientific Cluster (VSC).